\documentclass[preprint]{ptephy_v1}
\preprintnumber{KUNS-2622}
\usepackage{graphics}
\usepackage{braket}
\usepackage{bm}

\begin{document}
\title{Isospin Projected Antisymmetrized Molecular Dynamics and its Application to ${}^{10} \textrm{B}$}
\author{Hiroyuki Morita}
\author[1]{Yoshiko Kanada-En'yo}
\affil{Department of Physics, Kyoto University, Kyoto 606-8502, Japan. \email{morita.hiroyuki.47a@st.kyoto-u.ac.jp}}

\begin{abstract}
To investigate $pn$ pair correlations in $N=Z=\textrm{odd}$ nuclei, we  develop
a new framework based on the generator coordinate method of the 
$\beta\gamma$ constraint antisymmetrized molecular dynamics. 
In the framework, the isospin projection is performed before the energy variation to 
obtain the wave function optimized for each isospin. 
We apply the method to ${}^{10} \textrm{B}$ and show that it 
works well to describe coexistence of $T=0$ and $T=1$ states in low-energy 
spectra.  Structures of low-lying states and $pn$ correlations are investigated. 
Strong $M1$($0^+_1\rightarrow 1^+_1$) and 
$E2$($1^+_1\rightarrow 1^+_2$) transitions are understood by
the spin excitation of the $pn$ pair and the rotation of a deformed core, respectively.
\end{abstract}

\subjectindex{D11}
\maketitle
\section{Introduction}
Proton and neutron ($pn$) correlation is one of the key phenomena 
to understand properties of nuclei along the $N=Z$ line in the nuclear chart (see Ref. \cite{PN_rev} and references therein).
Unlike in identical pair correlations, two channels, $T=0$ and $T=1$, 
are possible in $pn$ pair correlations, and both channels play an important role in various nuclear structures. 
Competition between $T=0$ and $T=1$ $pn$ pairs 
has been attracting a great interest and discussed to describe 
level ordering of $J^\pi T=1^+0$ and $J^\pi T=0^+1$ states in 
$N=Z=\textrm{odd}$ nuclei and neighboring nuclei.
In the heavy-mass region, the competition has been investigated with 
mean-field approaches \cite{Gez_Mixed_Pairing,Cederwall_pn,Qi_pn,HFB_pn_1,HFB_pn_2,Yoshida_pn_1}.
In the light-mass region,
properties of a $pn$ pair at the nuclear surface have been studied in detail 
to understand low-lying spectra of $N=Z=\textrm{odd}$ nuclei \cite{Sagawa_three_body,Sagawa_pf_shell,Fujita_Sc}. 

As discussed in the study with a three-body model calculation \cite{Sagawa_three_body}, 
a proton and a neutron around a core nucleus
form a $pn$ pair in the $T=0$ or $T=1$ channel 
because of the $S$-wave attraction between nucleons.
The $T=1$ $pn$ pair is the mirror state of the dineutron pair, 
which is often discussed in neutron-rich nuclei.
The appearance of the $T=0$ $pn$ pair is peculiar to 
$N=Z=\textrm{odd}$ nuclei. Unlike the $T=1$ pair, the $T=0$ 
$pn$ pair has the finite intrinsic spin $S=1$ like a deuteron, 
and therefore it provides different $J$ states in low-energy region 
because of angular momentum coupling.
Indeed, it is experimentally known that many $T=0$ states 
coexist along with the $J^\pi T=0^+1$ state in low-energy spectra of 
$N=Z=\textrm{odd}$ nuclei \cite{PN_rev}. 
Moreover, a high $J$ state with $T=0$ comes down 
to the ground state in many $N=Z=\textrm{odd}$ nuclei in the light-mass region
except for those with closed-shell cores.
For example, the ground state spin parity of ${}^{10} \textrm{B}$ with $Z=N=5$ is 
$3^+$, for which the importance of three-nucleon forces is discussed in the 
no-core shell model calculations \cite{Navratil_nocore1,Navratil_nocore2}.

In systematic study of general $N=Z=\textrm{odd}$ nuclei, 
it is important to describe the competition between the $T=0$ and $T=1$ $pn$ 
correlations while taking into account spin configurations 
as well as nuclear deformation.
In the light-mass $N=Z$ region, cluster structure is another important feature 
which brings rich structures together with the $pn$ correlations.
The $pn$ pair feature and its dynamics can be affected by the cluster structure. 

To deal with these problems, we developed a new method 
based on the antisymmetrized molecular dynamics (AMD)\cite{Ono_Horiuchi_AMD,KanadaEnyo:1995tb,AMDsupp,AMD-ref,KanadaEn'yo:2012bj}
with constraint on quadrupole deformation parameters called 
$\beta\gamma$ constraint AMD \cite{Kimura_beta_GCM,Suhara_beta_gamma_GCM}.
The $\beta\gamma$ constraint AMD has been proved to be a useful approach to
study structures of stable and unstable nuclei. 
In the AMD method, 
existence of a core nucleus, a $pn$ pair, and a cluster structure are not {\it apriori}
assumed, but degrees of freedom of all nucleons are independently treated
in basis wave functions given by Slater determinants of Gaussian wave packets.
Nevertheless, if a system favors a structure with a $pn$ pair and cluster structures, such the structure is obtained in the energy variation.
In the new method, the isospin projection is performed 
before the energy variation to obtain the AMD wave function 
optimized for each isospin state. We call this method, 
the isospin projected $\beta\gamma$ constraint AMD, $T\beta\gamma$-AMD.
Energy levels of $N=Z=\textrm{odd}$ nuclei are calculated 
with the $T\beta\gamma$-AMD method combined with the generator coordinate method (GCM),
called $T\beta\gamma$-AMD+GCM.
To test the applicability of the new framework, we apply 
the method to $^{10}$B, in which $2\alpha+pn$ structures are found in low-lying $T=0$ and $T=1$ 
states. We investigate $pn$ correlations around the $2\alpha$ core, and 
show the importance of the finite spin of the $T=0$ $pn$ pair, which couples with 
the $pn$ motion and the core rotation, in the low-lying states of $^{10}$B.

The paper is organized as follows. We explain the framework of the $T\beta\gamma$-AMD
in the section \ref{formalism}, and the adopted effective nuclear 
interactions in section \ref{hamiltonian}.
We show the calculated results of $^{10}$B in section \ref{results}, and give 
discussion of structures of ${}^{10} \textrm{B}$ focusing on the $pn$ correlation in
section \ref{discussion}. Finally, a summary and an outlook are given in section \ref{summary}.

\section{Framework of $T\beta\gamma$ constraint AMD}
\label{formalism}

The antisymmetrized molecular dynamics is one of the useful approaches
to describe the cluster aspect of light nuclei. However, the conventional AMD method is not sufficient 
for $N=Z=\textrm{odd}$ nuclei, in which the isospin 
$T=0$ and $T=1$ states degenerate in the low-energy region.
In this section, we explain the framework of the isospin projected AMD, 
which is a newly developed method for the study of  $N=Z=\textrm{odd}$ nuclei.
We also describe the detailed formulation of the present calculation which is 
based on the isospin projected version of the $\beta\gamma$ constraint AMD
combined with the GCM.

\subsection{AMD framework}
The AMD framework is based on a variational method.
An AMD wave function is a Slater determinant of Gaussian wave packets:
\begin{equation}
\Ket{\Phi} = \mathcal{A}\Ket{\phi_1}\Ket{\phi_2}\cdots\Ket{\phi_A},
\end{equation}
where $\mathcal{A}$ refers to the antisymmetrization operator and ${\phi_i}$ refers to the $i$th single-particle wave function.
${\phi_i}$ is expressed by a product of the spatial$(\psi_i)$, spin$(\xi_i)$, and isospin($n_i$) parts as,
\begin{equation}
\Ket{\phi_i} = \Ket{\psi_i}\Ket{\xi_i}\Ket{n_i},
\end{equation}
where
\begin{equation}
\Braket{\bm{r}|\psi_i} = \left(\frac{2\nu}{\pi}\right)^{\frac{3}{4}}\exp\left[-\nu\left(\bm{r}-\frac{\bm{Z}_i}{\sqrt{\nu}}\right)^2+\frac{1}{2}\bm{Z}_i^2\right],
\end{equation}
\begin{equation}
\Ket{\xi_i} = \xi_{i\uparrow}\Ket{\uparrow}+\xi_{i\downarrow}\Ket{\downarrow},
\end{equation}
\begin{equation}
\Ket{n_i} = \Ket{p} \textrm{or} \Ket{n}.
\end{equation}
The parameters of Gaussian centroids 
$\left\{\bm{Z}_i\right\}_{i=1,2,\ldots,A}$ and those of spin
orientations $\left\{\bm{\xi}_i\right\}_{i=1,2,\ldots,A}$ are optimized in the energy 
variation, whereas each 
isospin configuration is fixed to be a isospin eigenstate as $\Ket{p}$ or $\Ket{n}$.

In the AMD wave function expressed by a Slater determinant, 
some symmetries such as the parity and rotational symmetries are usually broken.
To obtain physical wave functions for energy levels,
the parity and angular momentum 
projections are performed to restore the broken symmetries in the AMD framework.
In most cases, the parity projection is performed before the energy variation but 
the angular momentum projection is done after the variation 
to save computational costs except for 
the extended AMD calculation 
with the variation after the parity and angular-momentum projections (AMD+VAP${}^{J}$) \cite{Enyo_VAP}.
Namely, energy variation is performed for the AMD wave function projected on to the parity eigenstate with the parity projection operator $P^\pi$ as
\begin{eqnarray}
	&&\delta\frac{\Braket{\Phi^\pi|H|\Phi^\pi}}{\Braket{\Phi^\pi|\Phi^\pi}} = 0, \\
 &&\Ket{\Phi^\pi} = P^\pi\Ket{\Phi}.
\end{eqnarray}
The expectation value of an observable $O$ is calculated 
by the parity and angular momentum eigenstates
$\Ket{\Phi^{\pi J}_{MK}}$ projected from the 
obtained AMD wave function as,
\begin{eqnarray}
	&&\Braket{O}=\frac{\Braket{\Phi^{\pi J}_{MK}|O|\Phi^{\pi J}_{MK}}}{\Braket{\Phi^{\pi J}_{MK}|\Phi^{\pi J}_{MK}}},\\
 &&\Ket{\Phi^{\pi J}_{MK}} = P^\pi P^J_{MK}\Ket{\Phi},
\end{eqnarray}
where $P^J_{MK}$ is the angular momentum projection operator.
For numerical calculation of the energy variation, 
we adopt the frictional cooling method, 
which is a gradient method.

To describe the ground and excited states, we superpose 
various AMD wave functions based on the GCM with the $\beta\gamma$ constraint AMD method \cite{Suhara_beta_gamma_GCM}, which is an efficient method to 
choose basis AMD wave functions for a multi-configuration calculation.
 
In the $\beta\gamma$ constraint AMD, the energy variation is performed under 
the constraint on quadrupole deformation parameters $\beta$ and $\gamma$. 
The method has been proved to be useful to describe spatially developed cluster states 
as well as deformed states.
To perform the energy variation with a constraint $\bar{C}\left[\Phi\right]=C_i$, we add a penalty term to 
the energy expectation value as,
\begin{equation}
\Braket{H}\rightarrow\Braket{H}+\eta\left(\bar{C}\left[\Phi\right]-C_i\right)^2,
\end{equation}
where $\eta$ is an enough large positive value. 
After the energy variation with the
penalty term, we obtain the optimized configuration $\Phi\left(C_i\right)$
for the energy minimum state in the AMD model space 
with the condition $\bar{C}\left[\Phi\right]=C_i$. 

We superpose the wave functions obtained for 
different constraint values $C_i$ based on the GCM method with the generator coordinate $\bar{C}\left[\Phi\right]$. In the $\beta\gamma$ constraint AMD, 
$C_i$ indicates the set of the quadrupole deformation parameters, $C_i=(\beta_i,\gamma_i)$, which specifies
the nuclear deformation. 
Then, the GCM wave function is given by the superposition of the 
parity and angular momentum projected 
wave functions obtained by the $\beta\gamma$ constraint AMD as,
\begin{equation}
\Ket{\Psi} = \sum_{i=1}^{i_\textrm{max}}\sum_{K=-J}^{J}g^{\pi J}_{iK} P^\pi P^J_{MK}\Ket{\Phi\left(C_i\right)},
\end{equation}
where $K$ refers to the intrinsic magnetic quantum number and $i_\textrm{max}$ is the number of superposed wavefunctions.
The coefficients $\left\{g^{\pi J}_{iK}\right\}$ 
are determined by the diagonalization of the Hamiltonian and norm matrices.

\subsection{isospin projection}
In $N=Z=\textrm{odd}$ nuclei, the isospin projection is necessary to describe
different isospin states $T=0, 1$, which degenerate in a 
low-energy region.
As done for the parity projection, the isospin projection enables us to obtain the better wave function optimized for each isospin.
The isospin projection is also important to obtain proper intrinsic 
spin structures because 
the isospin of a $pn$ pair restricts 
the intrinsic spin of the pair because of the Fermi statistics of two nucleons. 

The isospin projection operator is defined as
\begin{equation}
P^T_{T_MT_K} = \frac{2T+1}{8\pi^2}\int d\Omega D^{T*}_{T_MT_K}\left(\Omega\right)R^T\left(\Omega\right),
\end{equation}
where $D^{T}_{T_MT_K}$ is the Wigner's $D$ function and $R^T\left(\Omega\right)$
is the SU(2) rotation operator in the isospin space.
In the present calculation of $N=Z$ nuclei, we approximately perform the isospin 
projection with the following operator, 
\begin{equation}
P^T = 1+\pi^T P_{pn},
\end{equation}
where $\pi^T=\left(-1\right)^Z, -\left(-1\right)^Z$ for $T=0, 1$ respectively.
Here $P_{pn}$ is the operator which exchanges the isospin, $\textrm{proton} \leftrightarrow \textrm{neutron}$, of all nucleons in the AMD wave function:
\begin{equation}
P_{pn}\Ket{\Phi}=\mathcal{A}\Ket{\psi_1}\Ket{\xi_1}\Ket{-n_1}\Ket{\psi_2}\Ket{\xi_2}\Ket{-n_2}\cdots\Ket{\psi_A}\Ket{\xi_A}\Ket{-n_A},
\end{equation} 
where $\Ket{-p}\equiv\Ket{n}$ and $\Ket{-n}\equiv\Ket{p}$. 
For a $N=Z=\textrm{odd}$ system having a proton and a neutron 
with a core nucleus, the operator $P^T$ is a good approximation 
of the isospin projection operator provided that the core nucleus is an
approximately $T=0$ state. 

\subsection{isospin projected AMD}
In the isospin projected AMD framework, 
the isospin projection is performed before the energy variation.
In the present calculation, the method is used for the 
$\beta\gamma$ constraint AMD. Namely, the
energy variation is performed with the $\beta$ and $\gamma$ constraints
after the isospin and parity projections but before the 
angular momentum projection as
\begin{eqnarray}
&&\delta\frac{\Braket{\Phi^{T\pi}|H|\Phi^{T\pi}}}{\Braket{\Phi^{T\pi}|\Phi^{T\pi}}}=0,\\
&&\Ket{\Phi^{T\pi}}=P^T P^\pi \Ket{\Phi},
\end{eqnarray}
with the penalty term as
\begin{equation}
\Braket{H}\rightarrow\Braket{H}+
\eta\left[\left(\beta_i\cos\gamma_i-\bar{\beta}\cos\bar{\gamma}\right)^2+
\left(\beta_i\sin\gamma_i-\bar{\beta}\sin\bar{\gamma}\right)^2\right],
\end{equation}
where $\bar{\beta}$ and $\bar{\gamma}$ are defined as 
\begin{equation}
\bar{\beta}\cos\bar{\gamma} = \frac{\sqrt{5\pi}}{3}\frac{2\Braket{z^2}-\Braket{x^2}-\Braket{y^2}}{R^2},
\end{equation}
\begin{equation}
\bar{\beta}\sin\bar{\gamma} = \sqrt{\frac{5\pi}{3}}\frac{\Braket{x^2}-\Braket{y^2}}{R^2},
\end{equation}
\begin{equation}
R^2 = \frac{5}{3}\left(\Braket{x^2}+\Braket{y^2}+\Braket{z^2}\right).
\end{equation}
We call this method, the $T$-projected $\beta\gamma$-constraint AMD ($T\beta\gamma$-AMD).
In this method, $\beta$ and $\gamma$ are the parameters which control
the development of clusters including $pn$ pairs.
In the application to ${}^{10} \textrm{B}$ nucleus with $2\alpha+pn$ structures,
$\gamma$ plays a role of the parameter for spatial development of the $pn$ pair 
whereas $\beta$ is related to the $\alpha$-$\alpha$ distance, as shown later.
After the energy variation, we perform the GCM calculation by superposing 
the projected AMD wave functions with respect to the discretized generator coordinates 
$\beta_i$ and $\gamma_i$ as, 
\begin{equation}
\Ket{\Psi} = \sum_{i=1}^{i_\textrm{max}}\sum_{K=-J}^{J}g^{T \pi J}_{iK} P^T P^\pi P^J_{MK}\Ket{\Phi\left(\beta_i \gamma_i\right)}.
\label{finalwf}
\end{equation}

In the present calculation, we omit the isospin mixing in the wave functions.
Note that the isospin mixing by the Coulomb force can be easily taken into account in the present framework by superposing different $T$ states in the GCM calculation.

\section{Hamiltonian and parameters
}
\label{hamiltonian}

The Hamiltonian used in the present work is 
\begin{equation}
H = K - K_\textrm{cm} + V_\textrm{c} + V_\textrm{ls} + V_\textrm{Coulomb},
\end{equation}
where $K$, $K_\textrm{cm}$, $V_\textrm{c}$, $V_\textrm{ls}$, and $V_\textrm{Coulomb}$ are the 
kinetic energy, kinetic energy of the center of mass, central force, spin-orbit force, and Coulomb force, respectively. The Coulomb force is  
approximated by a 7-range Gaussian as done in Ref. \cite{OnoCoulomb}.
For the central force and the spin-orbit force, the following effective nuclear forces
are adopted as done in the previous work \cite{Enyo_Morita_Kobayashi}. 
The Volkov No.2 force \cite{Volkov_Force} is used for $V_\textrm{c}$ as,
\begin{equation}
V_\textrm{c} = \sum_{i<j}\sum_{k=1, 2}v_k\exp\left[-\left(\frac{\bm{r}_i-\bm{r}_j}{a_k}\right)^2\right]\left(W+BP_\sigma-HP_\tau-MP_\sigma P_\tau\right), 
\end{equation}
where $v_1 = -$60.65 MeV, $v_2 = $61.14 MeV, $a_1 = $1.80 fm and $a_2 = $1.01 fm.
As for the Wigner, Majorana, Bartlett, and Heisenberg parameters, we adopt the same parameters as those used in the previous work \cite{Enyo_Morita_Kobayashi}, $W = 1-M = $0.40, $M = $0.60,
and $B = H = $0.06. These parameters reproduce the $\alpha\alpha$ scattering phase shift. Bartlett and Heisenberg parameters were fitted to the relative energy 
between $T = 0$ and $T = 1$ states of the ${}^{10} \textrm{B}$ spectra in the AMD+VAP${}^{J}$ calculation.

For $V_\textrm{ls}$, the spin-orbit term of the Gaussian 3-range soft-core force (G3RS) \cite{G3RS_Force1,G3RS_Force2} is used as,
\begin{equation}
V_\textrm{ls} = \sum_{i<j}\sum_{k=1, 2}u_k\exp\left[-\left(\frac{\bm{r}_i-\bm{r}_j}{b_k}\right)^2\right]\frac{1+P_\sigma}{2}\frac{1+P_\tau}{2}\bm{l}_{ij}\cdot\bm{s}_{ij},
\end{equation}
\begin{equation}
\bm{l}_{ij} = \left(\bm{r}_i-\bm{r}_j\right)\times\frac{\bm{p}_i-\bm{p}_j}{2},
\end{equation}
\begin{equation}
\bm{s}_{ij} = \bm{s}_i + \bm{s}_j ,
\end{equation}
where $b_1 = $0.60 fm, $b_2 = $0.447 fm. The strengths used in the present 
calculation are 
$u_1 = -u_2 = $1300 MeV, which were fit to reproduce the ls splitting between $3/2^-$
and $1/2^-$ states in ${}^{9} \textrm{Be}$ in the AMD+VAP${}^{J}$ calculation \cite{Enyo_Morita_Kobayashi}.

\section{Results}
\label{results}
To test the applicability of the $T\beta\gamma$-AMD, we apply it
to ${}^{10} \textrm{B}$ and show advantages 
of this method.
${}^{10} \textrm{B}$ is an ideal benchmark nucleus for the following reasons.
Firstly,
${}^{10} \textrm{B}$ is a $N=Z$ nucleus, in which different isospin states appear in its low-energy region. Indeed, 
$J^\pi T=3^+0, 1^+0$ and $0^+1$ states are
known in the low-lying $^{10}$B spectra. 
In second, ${}^{10} \textrm{B}$ is considered to have the $2\alpha+pn$ structure. It means that ${}^{10} \textrm{B}$ can be a good example 
of a system having a $pn$ pair around a deformed core.
We investigate structures of low-lying $T=0$ and $T=1$ states 
in ${}^{10} \textrm{B}$ while focusing on the $pn$ correlation around the $2\alpha$ core.

\subsection{Energy variation with and without the isospin projection}

In the framework of the isospin projected AMD, we obtain the 
state optimized for each isospin $T=0,1$ because the energy variation is 
performed for the isospin projected AMD wave function as explained previously.
The variation after the isospin projection (VAP${}^{T}$) is essential 
to clarify the nuclear structure along the $N=Z$ line
on the nuclear chart
because different isospin states have almost same energy and compete each other
in $N=Z=\textrm{odd}$ nuclei.
This is an advantage of the $T\beta\gamma$-AMD superior to the usual $\beta\gamma$-AMD, in which the variation is performed without the isospin
projection.
In this subsection, we compare the energy surface on the $\beta\gamma$ plane obtained by
the T$\beta\gamma$-AMD with that of the $\beta\gamma$-AMD
and show the availability of the 
variation after the isospin projection in the present framework, 
T$\beta\gamma$-AMD.

\begin{figure}[!ht]
\centering\includegraphics[width=\hsize]{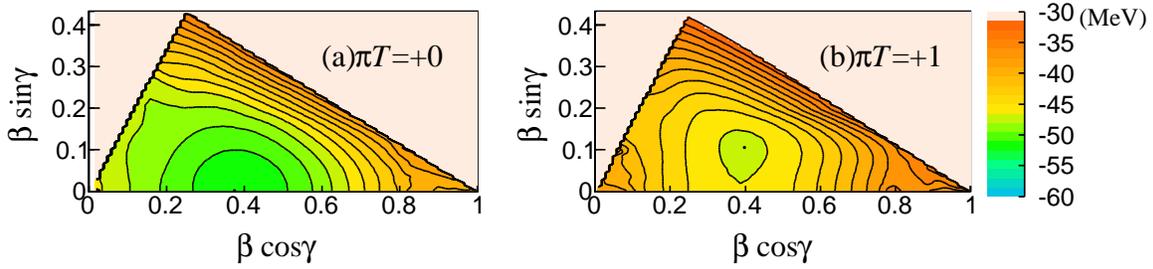}
\caption{
Energy surfaces 
on the $\beta\gamma$ plane of ${}^{10} \textrm{B}$ obtained by the
$T\beta\gamma$-AMD.
The left (right) panel shows the $T=0$($T=1$) $\pi=+$
energy surface.
The energy minimum of each energy surface is shown by a dot.
}
\label{figure1}
\end{figure}

First, we show the energy on the $\beta\gamma$ plane obtained by the $T\beta\gamma$-AMD.
Figure~\ref{figure1} shows the $T=0,1$ $\pi=+$
energy surfaces of ${}^{10} \textrm{B}$
obtained by the variation after the parity and isospin projections.
The minimum point in the $T=1$ plane (Fig.~\ref{figure1}(b)) is located at $\left(\beta\cos\gamma,\beta\sin\gamma\right)=\left(0.4,0.1\right)$ and
that in the $T=0$ plane (Fig.~\ref{figure1}(a)) is at $\left(0.38,0\right)$.
The finite $\gamma$ value is mainly caused by spatial development of the $pn$ pair
as discussed later.

\begin{figure}[!ht]
\centering\includegraphics[width=0.6\hsize]{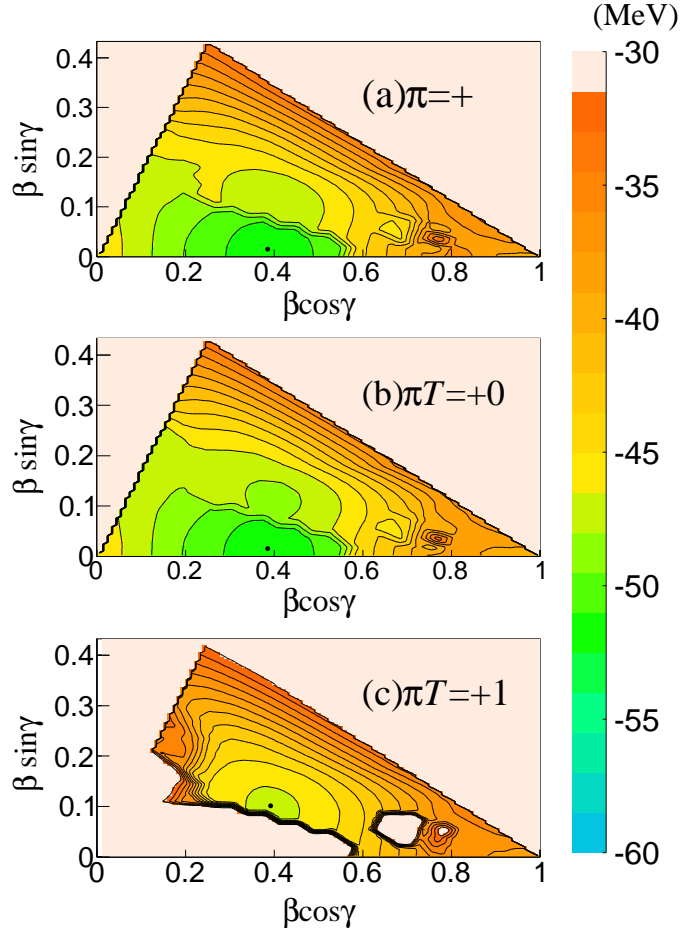}
\caption{
Energy surfaces 
on the $\beta\gamma$ plane of ${}^{10} \textrm{B}$ obtained by the VBP$^T$. 
(a) The $\pi=+$ energy surface 
obtained by the $\beta\gamma$-AMD variation without the isospin projection.
 (b) (c) The $T=0$ and $T=1$ projected $\pi=+$ energy surfaces of 
 the 
 VBP${}^{T}$.
The energy minimum of each energy surface is shown by a dot.
}
\label{figure2}
\end{figure}

Next, we discuss the results obtained by the $\beta\gamma$-AMD, 
which is a conventional AMD method with the parity projection without the isospin projection.
Figure~\ref{figure2}(a) shows the $\pi=+$ energy surface of ${}^{10} \textrm{B}$ obtained by the $\beta\gamma$-AMD.
It is notable that there is a gap along the line from (0.2,0.1) to (0.6,0).
This is caused by the drastic change of the intrinsic structure
from the small $\gamma$ region to the relatively large $\gamma$ region.

We also calculate the energy expectation values of the 
$T=0, 1$ eigenstates projected from the 
$\beta\gamma$-AMD wave functions obtained by the variation
before the isospin projection (VBP${}^{T}$).
The $T=0$ and $T=1$ $\pi=+$ energy surfaces of the VBP${}^{T}$ are shown in Figs. \ref{figure2}(b) and (c), respectively.
Also the $T=0, 1$ energy surfaces of the VBP${}^{T}$ show the discontinuity
along the same line on the $\beta\gamma$ plane  
originating in the difference in the intrinsic structure in two regions.
The $T=0$ energy surface in the small $\gamma$ region 
corresponds well to the $T=0$ result obtained by the VAP${}^{T}$ shown in 
Fig.~\ref{figure1}(a), whereas the $T=1$ energy surface in the large $\gamma$ region
is similar to the $T=1$ result of the VAP${}^{T}$ in Fig.~\ref{figure1}(b).
This means that, for the small $\gamma$ region, the VBP${}^{T}$ 
gives the wave functions almost same as those optimized for the $T=0$ component,
and for the large $\gamma$ region, it produces the solutions similar to those 
optimized for the $T=1$ component.

From these results, it is clear that the VBP${}^{T}$ method is not applicable to 
obtain the optimum wave functions on the $\beta\gamma$ plane for each isospin. 
As seen in Fig.~\ref{figure2}(b), the VBP${}^{T}$ fails to obtain the $T=1$ states in the small $\gamma$ region,
that is., the $T=1$ wave functions with prolate deformations.
As already shown in Fig.~\ref{figure1}, the $T=0$ and $T=1$ states 
exist in each point on the $\beta\gamma$ plane. However,  in the VBP${}^{T}$, 
the $T=0$ and $T=1$ states can not be optimized separately, but  
either of $T=0, 1$ states or a mixed state of $T=0$ and $T=1$ components 
is optimized.
The discontinuity on the $\beta\gamma$ plane in the VBP${}^{T}$ results
can be a crucial problem 
to describe the $\gamma$ mode, which is related to $pn$ pair motion, and to calculate excited states in each isospin channel.

On the other hand, the VAP${}^{T}$ technique enables us to optimize 
$T=0$ and $T=1$ components separately even though they almost degenerate 
at each constraint point. Consequently, the discontinuity
on the $\beta\gamma$ plane disappears and 
the smooth energy surfaces are obtained for each isospin states by the $T\beta\gamma$-AMD
as shown in Fig.~\ref{figure1}. This is one of the advantages of the present method
and important to describe the $\gamma$ mode for $pn$ pair motion
in detail. 

\subsection{Angular momentum projected energy surfaces and intrinsic structures}
\begin{figure}[!ht]
\centering\includegraphics[width=\hsize]{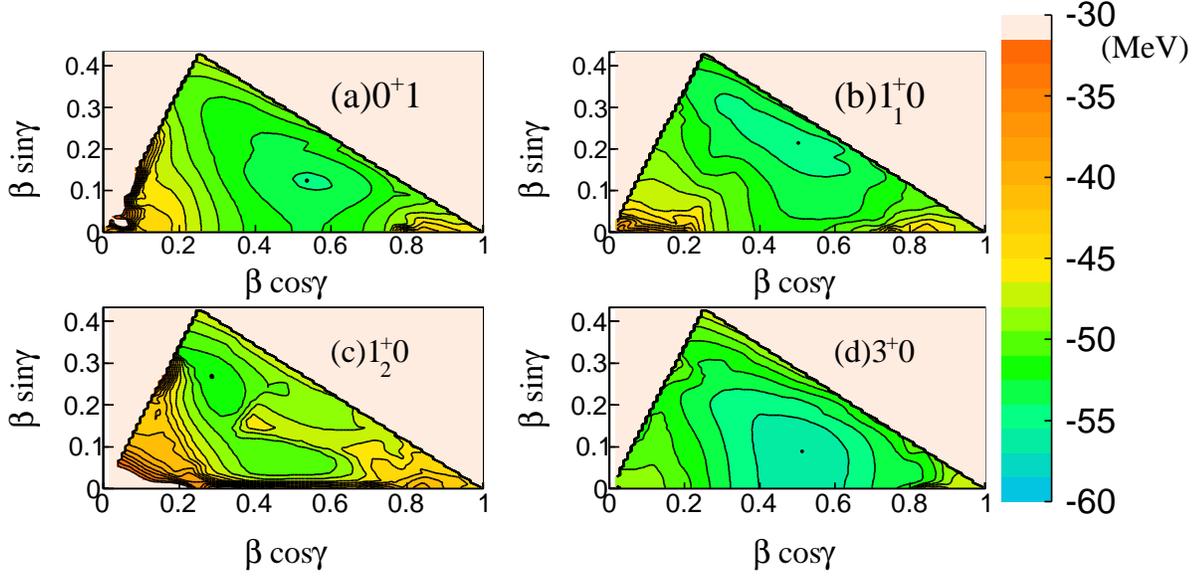}
\caption{$J^\pi$-projected energy surfaces of the VAP${}^{T}$
on the $\beta\gamma$ plane of ${}^{10} \textrm{B}$. 
(a), (b), (c) and (d) panels refer to the $0^+ 1$, $1^+_10$, $1^+_20$ and $3^+0$, respectively.
The energy minimum on each energy surface is shown by a dot.
}
\label{figure3}
\end{figure}

We discuss the angular momentum projected energies of the 
$T\beta\gamma$-AMD,
which are calculated by the angular momentum projection of the 
wave functions obtained by the $T\beta\gamma$-AMD.
Figure~\ref{figure3} shows the energy surfaces on the $\beta\gamma$ plane 
for  $J^\pi T=0^+1$, $3^+0$, and $1^+0$ states. 
Here the $K$-mixing is taken into account and the lowest energy
is shown at each point on the $\beta\gamma$ plane except for Fig.~\ref{figure3}(c). 
The $J^\pi T=1^+ 0$ states on the $\beta\gamma$ plane contain 
two components with different $K$ quanta corresponding to the low-lying $1^+_10$ and
$1^+_20$ states. Figures \ref{figure3}(b) and (c) show the energy surfaces for the 
first and the second lowest energies obtained by the $K$-mixing at each point.
The $3^+0$ energy surface is approximately described by the energy of the $K=3$ component 
on the $\beta\gamma$ plane, whereas
the $1^+_10$ and $1^+_20$ energy surfaces are  roughly described by the energies of 
the $K=1$ and $K=0$ components, respectively, 
though strictly speaking the $K$-mixing occurs for $\gamma\ne 0$ states.

Compared with the $T=0$ and $T=1$ 
energy surfaces before the angular momentum projection 
in Fig.~\ref{figure1}, 
the energy minimum positions are shifted toward 
larger deformation $\beta$ regions by the angular momentum projection
meaning that deformed states are more favored after the angular momentum projection.
The energy minimum state at $(\beta\cos\gamma,\beta\sin\gamma)=(0.54, 0.12)$
for $\beta=0.55, \gamma=12.5^\circ$
on the $J^\pi T=0^+ 1$ energy surface (Fig.~\ref{figure3}(a)) is 
the dominant component of the 
lowest isovector ($T=1$) state,  ${}^{10} \textrm{B}(0^+_1)$, 
and that at $\left(0.52, 0.08\right)$ for $\beta=0.53, \gamma=8.7^\circ$
on the $3^+0$ energy surface  (Fig.~\ref{figure3}(d)) approximately
describes the
lowest isoscalar ($T=0$) state, the ${}^{10} \textrm{B}$ 
ground state ($3^+_10$).
The energy minimum state at $\left(0.5,0.21\right)$ on the $J^\pi T=1^+_10$ energy 
surface and that at $(0.29,0.27)$ on the $J^\pi T=1^+_20$ energy surface
approximately correspond to 
the ${}^{10} \textrm{B}(1^+_1)$ and ${}^{10} \textrm{B}(1^+_2)$, respectively. 

\begin{figure}[!ht]
\centering\includegraphics[width=0.7\hsize,bb=0 0 566 481]{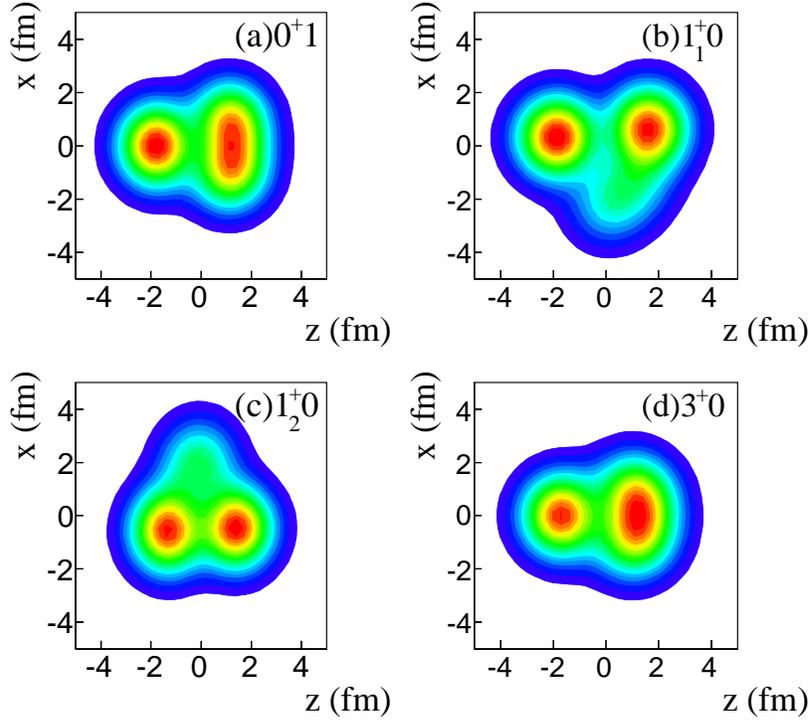}
\caption{
Intrinsic density of the states at the minimum points in the $J^\pi$-projected energy surfaces  of the VAP${}^{T}$.
Panels (a), (b), (c) and (d) show the density distribution projected to the xz plane of the $0^+ 1$, $1^+_10$, $1^+_20$ and $3^+0$, respectively.
}
\label{figure4}
\end{figure}

Let us discuss intrinsic structures of the states obtained by the $T\beta\gamma$-AMD.
The AMD wave functions before the projections at the energy minimums on the 
$J^\pi T=0^+1$, $3^+0$, $1^+_10$ and $1^+_20$ energy surfaces (Fig.~\ref{figure3}(a-d))
are regarded as the intrinsic states of the dominant components of the $0^+1$, $3^+0$, 
$1^+_10$, and $1^+_20$ states of $^{10}$B. 
In Fig.~\ref{figure4}, we show the intrinsic density distribution of the minimum energy states 
projected onto the $xz$ plane by integrating 
the density along the $y$-axis, $\rho\left(x, z\right) = \int \rho\left(x, y, z\right)dy$.
The density distribution shows the structure of 
a $2\alpha$ core with a $pn$ pair in the low-lying states of $^{10}$B.
In the $0^+1$ state, three $T=1$ $pn$ pairs with the anti-parallel spin configuration ($S=0$) 
are found in the right side forming an $\alpha$ cluster with a $S=0$ 
$pn$ pair nearby the $\alpha$ cluster.
In the $3^+0$ state, the $T=0$ ($S=1$) $pn$ pair is formed but it 
shows no spatial development because the $pn$ pair is rotating 
around the $2\alpha$ core in the $L=2$($D$-wave) and 
strongly attracted by the spin-orbit potential from the core.
In the $1^+_10$ and $1^+_20$ states,  
a $T=0$ $pn$ pair has the parallel spin configuration 
indicating the formation of the $T=0$, $S=1$ $pn$ pair.
The $pn$ pair in the $1^+0$ states is farther away from the $2\alpha$ core 
compared with those in the $0^+1$ and $3^+0$ states 
indicating remarkable spatial development of the $T=0$, $S=1$ $pn$ pair.

From the analysis of intrinsic structures 
obtained by the $T\beta\gamma$-AMD on the $\beta\gamma$ plane, 
it is found that, 
the $T=0,S=1$($T=1,S=0$) $pn$ pair is formed in the $T=0$($T=1$) states, 
and it goes away in the transverse direction from the $2\alpha$
with increase of $\gamma$. 
In other word, the $\gamma$ mode on the $\beta\gamma$ plane 
approximately corresponds the $pn$ pair motion relative to the $2\alpha$ core.

Let us come back to the energy surfaces shown in Fig.~\ref{figure3}.
The $0^+1$ energy surface shows a plateau from the minimum 
along the $\gamma$ parameter 
up to the $\gamma=60^\circ$ line, which corresponds to 
a soft mode of the
$T=1,S=0$ $pn$ pair motion relative to the $2\alpha$ core
in the $0^+_11$ state.
The $1^+_1 0$ energy surface shows a more remarkable plateau 
for the spatial development of the $T=0,S=1$ $pn$ pair.
In contrast, the $3^+0$ energy surface does not show such a plateau
toward the $\gamma=60^\circ$ direction
meaning that the $pn$ pair in the $3^+_10$ state is deeply bound near the core 
because of the spin-orbit potential from the core.

\subsection{GCM results}
We superpose $i_\textrm{max}=91$ wave functions on the $\beta\gamma$ plane for each isospin
and calculate energy spectra, moments, and transition strengths. The parity, isospin, and
the angular momentum projections with $K$-mixing are taken into account in the GCM 
calculation as described in Eq. (\ref{finalwf}).
The generator coordinates, $\beta$ and $\gamma$, effectively describe
the $pn$ pair motion relative to the $2\alpha$ core as well as the $\alpha$-$\alpha$ motion. 

\begin{figure}[!ht]
\centering\includegraphics[width=0.6\hsize,angle=270]{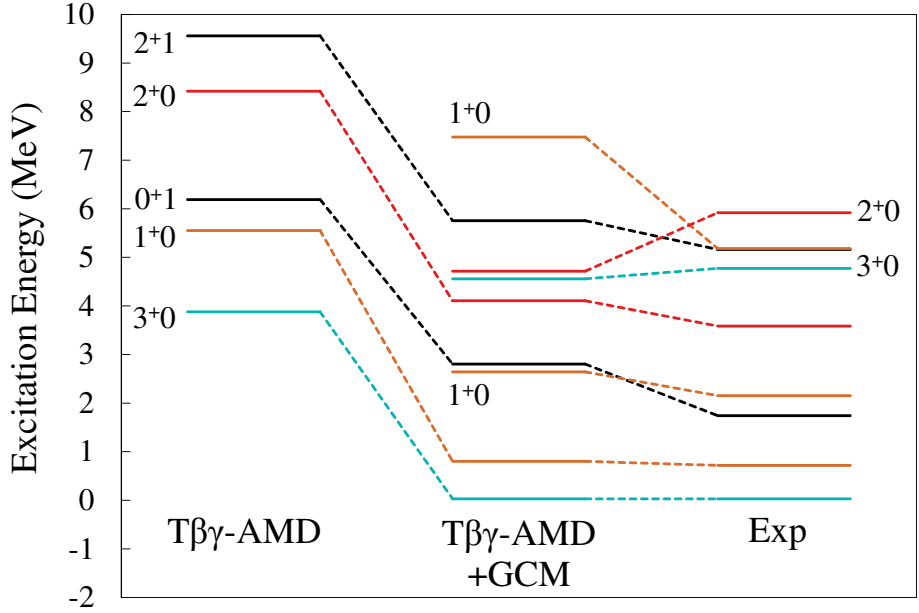}
\caption{
Spectra of ${}^{10} \textrm{B}$ calculated by the 
$T\beta\gamma$-AMD+GCM and those of the experimental data taken from Ref. \cite{ExpSpec1}.
The minimum energies in the $J^\pi$-projected energy surfaces of the 
$T\beta\gamma$-AMD measured from the $3^+_10$ energy of the $T\beta\gamma$-AMD+GCM are also shown.
}
\label{figure5}
\end{figure}

The $T=0$ and $T=1$ energy spectra of ${}^{10} \textrm{B}$ obtained by the GCM calculation are shown in Fig.~\ref{figure5}. 
Compared with the minimum energies of the $J^\pi T$ energy surfaces 
on the $\beta\gamma$ plane, about 4 MeV energy gain is obtained 
for the lowest $J^\pi T$ states by the superposition in the GCM calculation mainly 
because of the quantum mixing of spatial and spin configurations of the $pn$ pair. 
In particular, the $1^+_10$ states gains a larger energy
than the $3^+_10$ because of the spatial motion of the $pn$ pair  
along the plateau on the energy surface. As a result, the excitation energy 
of the $1^+_10$ state decreases in the GCM calculation. 
Moreover, we obtain excited $J^\pi T$ 
states as a result of configuration mixing as well as 
$K$-mixing in the GCM calculation.
The energy spectra of the GCM calculation reasonably reproduce the experimental spectra of $T=0$ and $T=1$ states.
Note that the relative energy between the $T=1$ states and $T=0$ states 
strongly depends on the interaction parameters, $B$ and $H$.

Table \ref{property} shows the calculated values of nuclear properties 
compared with the experimental data.
The present $T\beta\gamma$-AMD+GCM calculation
reproduces well the ground state properties such as the 
point-proton radius $r_p$, 
the electric quadrupole moment $Q$, and the magnetic moment $\mu$, 
and also describes reasonably the experimental data of $E2$ and $M1$ transition strengths.

\begin{table}[t]
\caption{
Nuclear properties and transition strengths of ${}^{10} \textrm{B}$ are shown.
The calculated values obtained by the $T\beta\gamma$-AMD+GCM and these obtained for the minimum energy states of the $J^\pi$-projected energy surfaces of the $T\beta\gamma$-AMD in Fig.~\ref{figure3} are shown.
The results of the AMD+VAP${}^{J}$ calculation in Ref. \cite{Enyo_Morita_Kobayashi} are also shown.
The experimental proton radii are derived from the charge radii in \cite{ProtonRadii}.
Other experimental data are taken from \cite{ExpSpec1,ExpSpec3}.
}
\label{property}
\begin{center}
\begin{tabular}{ccccc}
\hline
observable&$T\beta\gamma$-AMD&$T\beta\gamma$-AMD+GCM&AMD+VAP${}^{J}$&Exp\\
\hline\hline
$r_p\left(3_1^+0\right)$ ($\textrm{fm}$)&2.3&2.4&2.33&2.28(5)\\
$Q\left(3_1^+0\right)$ ($e$ $\textrm{fm}^2$)&7.5&8.4&8.2&8.47(6)\\
$\mu\left(3_1^+0\right) (\mu_N)$&1.8&1.8&1.85&1.8006\\
$\mu\left(1_1^+0\right) (\mu_N)$&0.8&0.8&0.84&0.63(12)\\
\hline
$B\left(E2;1_1^+0\rightarrow3_1^+0\right)$&3.0&4.0&3.6&4.14(2)\\
$B\left(E2;1_1^+0\rightarrow1_2^+0\right)$&&9.2&10.1&15.6(17)\\
$B\left(E2;1_2^+0\rightarrow3_1^+0\right)$&&2.0&1.3&1.7(2)\\
$B\left(E2;1_3^+0\rightarrow3_1^+0\right)$&&0.1&&\\
$B\left(E2;1_1^+0\rightarrow1_3^+0\right)$&&2.8&&\\
$B\left(E2;1_2^+0\rightarrow1_3^+0\right)$&&1.5&&\\
\hline
$B\left(M1;0_1^+1\rightarrow1_1^+0\right)$&8.7&15.0&14.7&7.5(32)\\
$B\left(M1;1_2^+0\rightarrow0_1^+1\right)$&&0.1&0.0&0.19(2)\\
$B\left(M1;0_1^+1\rightarrow1_3^+0\right)$&&0.0&&\\
\hline
\end{tabular}
\end{center}
\end{table}

\begin{figure}[!ht]
\centering\includegraphics[width=\hsize]{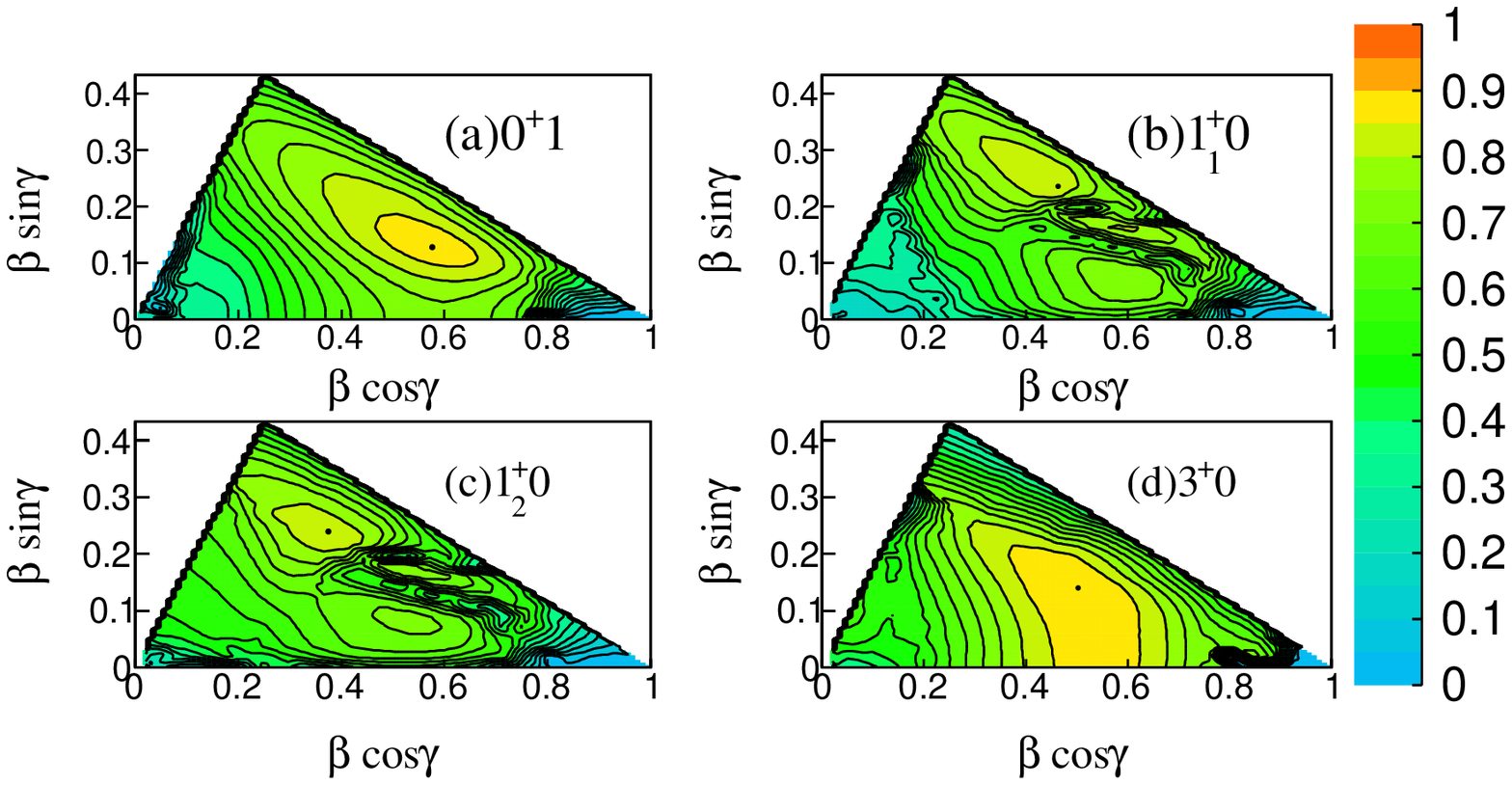}
\caption{Overlap amplitudes
on the $\beta\gamma$ plane of ${}^{10} \textrm{B}$. 
Panels (a), (b), (c) and (d) refer to the $0^+ 1$, $1^+_10$, $1^+_20$, and $3^+0$ states, respectively.
The maximum of each surface is shown by a dot.
}
\label{figure6}
\end{figure}

In Fig.~\ref{figure6}, we show the overlap amplitudes of the $0^+1$, $1_1^+0$, $1_2^+0$ and $3^+0$ states 
obtained by the GCM calculation with the basis wavefunctions $\Phi\left(\beta_i\gamma_i\right)$ on the $\beta\gamma$ plane.
The overlap amplitudes are calculated by the projection to the subspace composed of $J^\pi$-projected states, $\left\{P^T P^\pi P^J_{MK}\Ket{\Phi\left(\beta_i\gamma_i\right)}\right\}_{K=-J,\ldots,J}$.
For the ${}^{10} \textrm{B}(1^+_10)$ and ${}^{10} \textrm{B}(0^+_1)$ states,
the overlaps are distributed widely along the plateaus 
indicating the spatial development of the
$T=0, S=1$ and $T=1,S=0$ $pn$ pairs, respectively. 
For the ${}^{10} \textrm{B}(3^+_10)$, the overlap
is distributed in the small $\gamma$ region  
corresponding to the flat region around the energy minimum 
toward the $\gamma=0$ line on the $J^\pi T=3^+0$-projected  
energy surface (Fig.~\ref{figure3} (d)).
This result indicates that 
the ${}^{10} \textrm{B}(3^+_10)$ has less spatial development of 
the $pn$ pair and is regarded as the almost prolately deformed 
state with $\gamma$ fluctuation. 

\section{Discussion}
\label{discussion}
In this section, we discuss the structures of the low-lying states in ${}^{10} \textrm{B}$ 
and describe the spin and spatial configurations of the $pn$ pair.

\begin{figure}[!ht]
\centering\includegraphics[width=0.7\hsize,bb=0 0 560 700, angle=270]{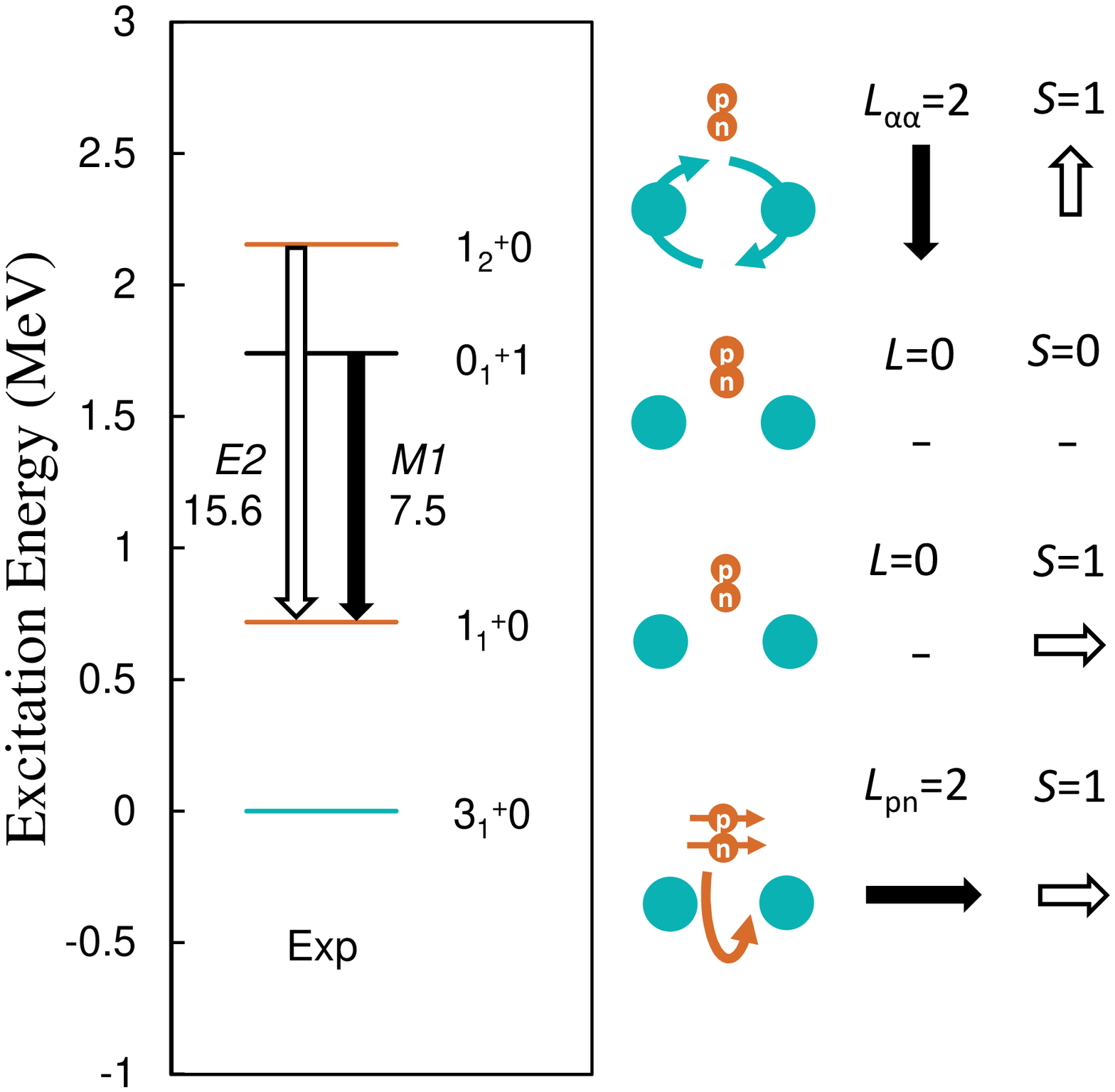}
\caption{
The schematic figure of spin configurations of the $pn$ pair and coupling with the orbital angular momentum in the low-lying states of ${}^{10} \textrm{B}$.
The experimental spectra are also shown.
$L$, $L_{\alpha\alpha}$ and $L_{pn}$ denote to the total orbital angular momentum, the orbital angular momentum of the $2\alpha$ core and that of the center of mass motion of the  $pn$ pair.
$S$ means to the intrinsic spin of the $pn$ pair.
}
\label{figure7}
\end{figure}

As discussed previously, 
the $T=0$ and $T=1$ states of  ${}^{10} \textrm{B}$
have dominantly the $S=1$ and $S=0$ $pn$ pairs 
around the $2\alpha$ core, respectively. 
The finite spin ($S=1$) of the $T=0$ pair couples with the orbital angular momentum
$L$ to the total angular momentum $J$. 
In the $2\alpha+pn$ structures, the orbital angular momentum ($L_{pn}$) for the 
$pn$ pair motion relative to the $2\alpha$ core and that ($L_{2\alpha}$)
for the $2\alpha$ core rotation contribute to the total orbital angular momentum $L$.
Table. \ref{spinexp} shows the expectation values $\Braket{\bm{L}^2}$ 
and $\Braket{\bm{S}^2}$ of the squared total orbital angular momentum 
and total spin angular momentum of the  $3^+_10$, $1^+_10$, $1^+_20$, and $0^+_11$ states obtained  by the $T\beta\gamma$-AMD+GCM. 
$\Braket{\bm{S}^2}=2.0$ of the $T=0$ states 
indicates the almost pure $S=1$ $pn$ pair 
in the $3^+_10$, $1^+_10$, and $1^+_20$ states,
and $\Braket{\bm{S}^2}=0.3$ of the $T=1$ state means the dominant 
$S=0$ component with slight mixing of the $S=1$ component in the  
$0^+_11$ state.
It should be commented that the present results of $\Braket{\bm{L}^2}$ and $\Braket{\bm{S}^2}$ are 
almost consistent with those of the no-core shell model calculation in Ref. \cite{pshell_nocore}, in which
analysis of the LS-coupling scheme was performed for light nuclei.

Then the structures of the $0^+1$, $3^+0$, $1_1^+0$ and $1_2^+0$ states are understood by the $T=0$ $S=1$ $pn$ pair
and $T=1$ $S=0$ $pn$ pair moving around the $2\alpha$ core as follows.
We show schematic figures of spin configurations of the $pn$ pairs and their coupling with orbital angular momentum in Fig. \ref{figure7}.
The $0^+1$ state has the dominant $L=0$ component with the  
$S=0$ $pn$ pair in the $S$-wave around the $2\alpha$ core.   
This state is the isobaric analogue state of the ${}^{10} \textrm{Be}$ having 
a $S=0$ $nn$ pair around the $2\alpha$ core. 
The $1_1^+0$ state dominantly contains the $L=0$ component 
with the $S=1$ $pn$ pair in $S$-wave.
This state is regarded as the spin partner of the $0_1^+1$ state. 
Because of the spin-flip transition in the $pn$ pair from $S=0$ to $S=1$, it 
has the strong $M1$ transition from the $0_1^+1$ to the $1_1^+0$,
as shown in Table \ref{property}.
The $1_2^+0$ state appears from the excitation of the orbital angular momentum 
from $L=0$ to $L=2$ mainly contributed by the rotation of the $2\alpha$ core
($L_{2\alpha}=2$). The strong $E2$ transition from the $1_2^+0$ state to 
the $1_1^+0$ state (see Table \ref{property}) shows the feature of the $L_{2\alpha}=2$ excitation in the $1_2^+0$ state.
The $3_1^+0$ state has the dominant $L=2$ component, which is mainly contributed 
by the orbital angular momentum $L_{pn}=2$ of the $S=0$ $pn$ pair.
From the cluster point of view, 
the $S=1$ $pn$ pair aligned to $L_{pn}=2$ feels strong spin-orbit potential from 
the $2\alpha$ core.
The strong attraction of the spin-orbit potential for the $S=1$ $pn$ pair in the $D$-wave
is the origin of the level inversion in ${}^{10} \textrm{B}$
between the $1_1^+0$ and $3_1^+0$ states with the $S$-wave 
and $D$-wave $pn$ pairs, respectively as pointed out in Ref. \cite{Enyo_Morita_Kobayashi}.
Note that, it corresponds to the $p_{3/2}^2$ configuration in the $jj$-coupling scheme.

\begin{table}[t]
\caption{
The expectation values $\Braket{\bm{L}^2}$ and $\Braket{\bm{S}^2}$ of the squared total orbital angular momentum and total spin angular momentum.
The values obtained by the $T\beta\gamma$-AMD+GCM and those calculated for the energy minimum states of the $J^\pi$-projected energy surfaces in Fig. \ref{figure3}($T\beta\gamma$-AMD).
}
\label{spinexp}
\begin{center}
\begin{tabular}{ccccc}
\hline
state&\multicolumn{2}{c}{$\Braket{\bm{L}^2}$}&\multicolumn{2}{c}{$\Braket{\bm{S}^2}$}\\
\hline
&$T\beta\gamma$-AMD&$T\beta\gamma$-AMD+GCM&$T\beta\gamma$-AMD&$T\beta\gamma$-AMD+GCM\\
\hline\hline
$3_1^+0$&7.1&7.2&2.0&2.0\\
$1_1^+0$&1.0&0.4&2.0&1.9\\
$1_2^+0$&&5.4&&1.9\\
$0_1^+1$&0.3&0.3&0.3&0.3\\
\hline
\end{tabular}
\end{center}
\end{table}

\section{Summary and outlook}
\label{summary}
We developed a new framework of the isospin projected AMD 
with the $\beta\gamma$ constraint and the GCM
called $T\beta\gamma$-AMD+GCM for study of 
$N=Z=\textrm{odd}$ nuclei. To test the applicability of the method we applied it to ${}^{10} \textrm{B}$.
The formation of the $S=1$ and $S=0$ $pn$ pairs around the $2\alpha$ core 
in the low-lying $T=0$ and $T=1$ states of ${}^{10} \textrm{B}$ is described 
in the $T\beta\gamma$-AMD+GCM calculation.
The spatial development of the $T=0$ and $T=1$ $pn$ pairs 
as well as the core deformation is controlled
by the $\beta\gamma$ constraint in the isospin projected AMD. 
By superposition of the optimized wave functions on the $\beta\gamma$ plane 
with the GCM, the spatial motion of the $pn$ pair as well as the 
core rotation is taken into account.

The $T\beta\gamma$-AMD+GCM calculation reproduces reasonably 
the properties of low-lying states of ${}^{10} \textrm{B}$.
The structures of the lowest four states 
($3^+_10$, $1^+_10$, $1^+_20$ and $0^+_11$) are understood by the 
angular momentum coupling of the $pn$ pair internal spin ($S$), 
its spatial motion, and the core rotation. 
The $3^+_10$($1^+_10$) state is described by the 
$2\alpha+pn$ structure having the $T=0,S=1$ $pn$ pair
in the $D$-wave($S$-wave) around the $2\alpha$ core.
Because of the spin-orbit attraction from the core 
for the $S=1$ $pn$ pair in the $D$-wave, the $3^+_10$ state 
comes down to the ground state.
The $1^+_20$ appears from the excitation of the $2\alpha$ core rotation
and has the strong $E2$ transition to the $1^+_10$ state. 
The $0^+_11$ state, which is the isobaric analogue state of the $^{10}$Be ground state,
is regarded as the spin partner of the $1^+_10$ state, and it has the strong $M1$ transition from the  $1^+_10$ state.

It was found that the $T\beta\gamma$-AMD+GCM can describe 
spin configurations and spatial development of the $pn$ pair in $T=0$ and $T=1$ states
as well as the core deformation and rotation. 
The $pn$ pair motion relative to the core and also the shape fluctuation of the deformed core are taken into account by the $\beta\gamma$ constraint with the GCM.
This is one of the advantages supplier to the AMD+VAP$^J$ which is based on the
the $J^\pi$-projected wave function of a single Slater determinant.
It was also shown that the isospin projection before the energy variation is necessary to obtain the optimum solution for each isospin state in $Z=N=\textrm{odd}$ nuclei, in which different isospin states almost degenerate in the low-energy region.

The present method is expected to be useful to investigate structure with 
a $pn$ pair around a deformed core in other $N=Z=\textrm{odd}$ such as 
$^{22} \textrm{Na}$ and $^{26} \textrm{Al}$, 
in which various $J^\pi T$ states appear in low-lying spectra. 
In principle, in the $T\beta\gamma$-AMD+GCM, existence of a $pn$ pair nor clusters
is not \textit{apriori} assumed. Therefore, it is applicable to general 
$N=Z=\textrm{odd}$ nuclei, and might enable us to make systematic study of 
$pn$ pair correlations in nuclei along the $N=Z$ line.
 
\section*{Acknowledgment}
The author would like to thank Dr.~Kobayashi for fruitful discussions.
The computational calculations of this work were performed by using the
supercomputer in the Yukawa Institute for theoretical physics, Kyoto University. 
This work was supported by 
JSPS KAKENHI Grant Number 26400270.

\end{document}